# The quantum state as spatial displacement

**Peter Holland**
Green Templeton College
University of Oxford
Oxford OX2 6HG
England

peter.holland@gtc.ox.ac.uk

**Abstract**: We give a simple demonstration that the Schrödinger equation may be recast as a self-contained second-order Newtonian law for a congruence of spacetime trajectories. This provides a pictorial representation of the quantum state as the displacement function of the collective whereby quantum evolution is represented as the deterministic unfolding of a continuous coordinate transformation. Introducing gauge potentials for the density and current density it is shown that the wave-mechanical and trajectory pictures are connected by a canonical transformation. The canonical trajectory theory is shown to provide an alternative basis for the quantum operator calculus and the issue of the observability of the quantum state is examined within this context. The construction illuminates some of the problems involved in connecting the quantum and classical descriptions.

### 1. Introduction

An unfortunate by-product of the historical debate on the interpretation of quantum mechanics is that physical ideas that may have informed the development of the subject have been marshalled into the siding of 'mere philosophy'. This has been the fate of the spacetime trajectory picture of quantum evolution, which is still widely assumed to be associated just with an interpretation of the theory, i.e., to depend on optional assumptions that are not inherent in the scheme of ideas that is generally accepted as constituting 'quantum theory'. In fact, it is an ineluctable mathematical property that the conception of a physical state based on the deterministic spacetime trajectory – comprising simultaneously well-defined position and momentum variables at each point – is implicit in the quantum description *whatever the interpretation*. More precisely, a self-contained theory of trajectories with second-order Newton-style dynamics may be obtained from the first-order Schrödinger equation by a change of variables [1,2]. In this formulation the quantum state is represented by the displacement function of a continuum of interacting 'particles', which involves a congruence rather than a single trajectory because wave mechanics is a field theory. The characteristic features of the wavefunction version of state are represented by distinctive properties of the congruence. For example, the unitary evolution of the wavefunction corresponds to the deterministic unfolding of a continuous coordinate transformation and the single-valuedness of the wavefunction to non-crossing of the trajectories. Indeed, one may make the displacement function of the collective the basis of the quantum description with the wavefunction being regarded as a derived quantity.

The trajectory conception of state emerged following many years of studies connected with the de Broglie-Bohm theory [3] and related computational work [4] and was first clearly elaborated by the author [1,2]. If the wavefunction is represented by hydrodynamic variables [5], the hydrodynamic and trajectory notions of state stand in relation to one another as the Eulerian and Lagrangian pictures of fluid dynamics [6] and we draw extensively upon the terminology and methods of that discipline. Our reference to 'particles' is to be construed as referring to fluid elements but no interpretative commitment to this imagery is required (they may be considered as 'elements of probability'). It is emphasized that the trajectory model is not a hidden-variable theory; it arises simply from a transformation of the independent and dependent variables employed in wave mechanics. A hidden-variable theory in this context would endow one of the trajectories in the congruence composing the state with some special property, be it a particular label or additional structure such as a corpuscle (as in the de Broglie-Bohm theory [3]), with the aim of explaining the quantum statistical predictions as the outcome of well-defined and causally connected individual events. Indeed, the trajectory representation of the state is particularly suited to describing how individuals may make up a statistical ensemble, something that is difficult to achieve using the wavefunction, but the additional assumptions required to do this are not part of the theory presented here. It has scarcely been noticed that much of the literature devoted to trajectory theories has been misleadingly classified as pertaining to 'interpretation' when in reality many of the results relate to quantum theory itself, albeit in this unfamiliar representation. A corollary is that problems ascribed to, say, the de Broglie-Bohm theory may in fact be issues to do with quantum mechanics that would be expressed in different terms (if at all) in the usual wavefunction approach.

Our purpose here is to provide an introductory account of the trajectory conception of quantum dynamics and investigate some of its key features. In Sec. 2 we present a simple derivation of the second-order Newton-like version of Schrödinger's equation that is more transparent than the demonstration in [1]. A particular aim is to exhibit the intimate relation between the wave-mechanical and trajectory pictures by showing how they are connected by a canonical transformation when each is expressed in suitable phase space coordinates. To this end, in Sec. 3 we first recall the usual Hamiltonian approach to quantum mechanics. We then develop in Sec. 4 a novel phase space formulation of quantum mechanics in its Eulerian hydrodynamical form that is more suited to our needs by introducing potentials for the density and current density. The potentials obey a second-order field equation and exhibit a gauge freedom that may be exploited to simplify the theory. This second-order representation of the Schrödinger equation is somewhat analogous to writing Maxwell's equations in terms of the electromagnetic potentials. In Sec. 5 we set up the desired canonical transformation linking the potentials-based canonical theory with a canonical formulation of the trajectory model in which the potentials transform into the displacement function. A significant point is that the potentials in the latter representation are gauge invariant quantities. In this endeavour we follow and develop a method introduced previously in classical fluid mechanics to connect the Eulerian and Lagrangian pictures [7], a procedure that is superior in several respects to the quantum canonical formulation used previously [1]. In Sec. 6 it is shown how the phase space trajectory theory may be used to represent the quantum operator algebra and in Sec. 7 we address the problem of the empirical determination of the state as it arises in the hydrodynamic formulation. Sec. 8 presents comments on how the trajectory view may provide the basis of a common language for the quantum and classical descriptions and how it thereby affords insight into the problems that arise in seeking to connect them.

## 2. Schrödinger's equation as Newton's law for a continuum of particles

### *2.1 Eulerian quantum hydrodynamics*

We are going to show how the Schrödinger equation,

$$i\hbar \frac{\partial \psi}{\partial t} = -\frac{\hbar^2}{2m} \frac{\partial^2 \psi}{\partial x_i \partial x_i} + V\psi, \qquad (2.1)$$

may be rewritten as Newton's law for a cloud of interacting particles pursuing spacetime trajectories. To this end we first rewrite it as two coupled real equations using the polar representation of the wavefunction, $\psi = \sqrt{\rho} e^{iS/\hbar}$:

$$\frac{\partial \rho}{\partial t} + \frac{\partial}{\partial x_i}\left(\rho \frac{1}{m} \frac{\partial S}{\partial x_i}\right) = 0 \qquad (2.2)$$

$$\frac{\partial S}{\partial t} + \frac{1}{2m} \frac{\partial S}{\partial x_i} \frac{\partial S}{\partial x_i} + V_Q + V = 0 \qquad (2.3)$$

where

$$V_Q(x) = -\frac{\hbar^2}{2m\sqrt{\rho}} \frac{\partial^2 \sqrt{\rho}}{\partial x_i \partial x_i} \qquad (2.4)$$

is the quantum potential and $i,j,k,\ldots = 1,2,3$. These equations are derived though multiplication and division by $\psi$ and hence hold in non-nodal ($\psi \neq 0$) regions. The field variables $\rho, S$ - which now represent the quantum state - inherit the continuity, boundary and single-valuedness conditions obeyed by $\psi$. The latter condition is expressed by the quantization condition

$$\oint \frac{\partial S}{\partial x_i} dx_i = nh, \quad n \in \mathbb{Z}, \qquad (2.5)$$

where the integration is over a loop fixed in space along which $\rho \neq 0$. The singularities in the phase $S$ occur at nodes.

This way of articulating the Schrödinger equation is the basis of its Eulerian-picture hydrodynamic representation [5] where the fluid functions are expressed with respect to a fixed system of coordinates. Thus, (2.3) may be regarded as a Bernoulli-type equation with $\rho$ the number density and $S$ the velocity potential of a putative continuous 'quantum fluid'. Writing

$$v_i(x,t) = \frac{1}{m}\frac{\partial S(x,t)}{\partial x_i} \tag{2.6}$$

for the velocity field we obtain from (2.2) the fluid continuity equation and, differentiating (2.3) with respect to $x$, an Euler-type force law:

$$\frac{\partial \rho}{\partial t} + \frac{\partial}{\partial x_i}(\rho v_i) = 0 \tag{2.7}$$

$$\frac{\partial v_i}{\partial t} + v_j \frac{\partial v_i}{\partial x_j} = -\frac{1}{m}\frac{\partial}{\partial x_i}(V + V_Q). \tag{2.8}$$

Following (2.5), the circulation obeys $\oint v_i \, dx_i = nh/m$ so the velocity field is single-valued and irrotational except along nodal lines where it is singular:

$$\varepsilon_{ijk} \frac{\partial v_k}{\partial x_j} = \sum_\mu n_\mu h \int_{L_\mu} \delta(x - x_\mu) dx_{\mu i} \tag{2.9}$$

with $x_{\mu i}$ the coordinates of the $\mu$th nodal line $L_\mu$ and $\varepsilon_{ijk}$ the antisymmetric symbol with $\varepsilon_{123} = 1$. The fluid therefore possesses quantized vortices [8,9]. From a mathematical perspective an advantage of the hydrodynamic formulation is that it involves only quantities $\rho, v_i$ that are independent of the irrelevant global phase. A disadvantage is that the linearity of quantum mechanics is rendered somewhat awkwardly in these variables although they do make interference effects transparent.

If we adopt equations (2.7) and (2.8) as the quantum evolution equations they must be supplemented by the subsidiary condition (2.6) subject to (2.5). To show that we indeed obtain (2.1) from the hydrodynamic equations, we note that (2.8) with (2.6) inserted implies the equation

$$\frac{\partial}{\partial x_i}\left(\frac{\partial S}{\partial t} + \frac{1}{2m}\frac{\partial S}{\partial x_j}\frac{\partial S}{\partial x_j} + V_Q + V\right) = 0. \tag{2.10}$$

It follows that the quantity in brackets is an arbitrary function of $t$ and, absorbing this function into a redefined $S$, we obtain (2.3). Combining the latter with (2.7), (2.1) follows. Clearly, if (2.6) is not imposed as a subsidiary condition the flow implied by (2.7) and (2.8) will be more general than that described by the Schrödinger equation. In fact, in the more general case the equations can still be combined into Schrödinger form but the phase, $\int mv_i(x)dx_i$, becomes path dependent and the 'wavefunction' multivalued.

### 2.2 Preservation of initial conditions

An important aspect of the derivation of equation (2.1) from the hydrodynamic equations is that the subsidiary condition (2.6) need be assumed only at one instant. Hence, taking the latter to be $t = 0$, the condition represents a constraint just on the initial velocity field:

$$v_{0i}(x) = m^{-1}\partial_i S_0(x). \tag{2.11}$$

To prove this, we start by subtracting the Euler equation (2.8) where $v_i = m^{-1}\partial_i S$ from that equation when $v_i$ is arbitrary:

$$\left(\frac{\partial}{\partial t} + v_i \partial_i\right)(mv_i - \partial_i S) = -\frac{1}{m}(mv_j - \partial_j S)\partial_{ij} S. \tag{2.12}$$

To show that $mv_i = \partial_i S$ is the unique solution to this equation for all $t$ given (2.11), we use the method of characteristics. This entails passing to moving coordinates $x_i = q_i(t)$ defined by the integral curves of the velocity, obtained by solving the differential equation $v_i(x_i = q_i(t), t) = \dot{q}_i$. Then, evaluating (2.12) in these coordinates, we obtain

$$\frac{d}{dt}\left(m\dot{q}_i - \partial_i S\big|_{x=q(t)}\right) = -\frac{1}{m}\left(m\dot{q}_j - \partial_j S\big|_{x=q(t)}\right)\partial_{ij} S\big|_{x=q(t)} \tag{2.13}$$

where $d/dt = \partial/\partial t + \dot{q}_i \partial_i$. This relation has the form of a first-order linear ordinary differential equation $\dot{X}_i = A_{ij}(t) X_j(t)$ for which continuity of the matrix $A_{ij}(t)$ guarantees the existence and uniqueness of solutions $X_i(t)$ [10]. Then, since $m\dot{q}_i - \partial_i S = 0$ is a solution of (2.13), this is the unique solution for all $t$ if it holds at $t = 0$, granted the continuity of the function $\partial_{ij} S(x = q(t))$. Returning to space coordinates, we have proved that (2.11) implies (2.6) for all $t$.

The preservation of gradient flow is a classical result and the details of the quantum contribution to the force in Euler's law are unimportant, other than that it contributes to an acceleration potential. There are a variety of alternative proofs, such as employing Kelvin's circulation theorem or Cauchy's vorticity theorem (see [11] for a compendium of methods). If we wish to characterize the quantum state by the $\rho, v_i$ variables the initial condition (2.11) is understood.

The theorem just proved concerns a condition for the identity of two flows corresponding to the same density and relies on Euler's force equation. In parenthesis, we remark that a complementary result holds for the density: given a flow $v_i$ and two possible densities $\rho, |\psi|^2$ obeying the continuity equation, the initial constraint $\rho_0 = |\psi_0|^2$ ensures $\rho = |\psi|^2$ for all $t$. To prove this we may again employ the characteristics but now appeal to the continuity equation. Thus, writing $f(x,t) = \rho/|\psi|^2$ we deduce from (2.7) that $df(x = q(t), t)/dt = 0$ or $f(x = q(t), t) = \text{const}$. Then, if $\rho_0(x) = |\psi_0(x)|^2$ for each $x$, we have $f = 1$ for all $x$ and returning to space coordinates $\rho(x,t) = |\psi(x,t)|^2$ for all $t$.

The characteristics may also be used to show that the quantization condition (2.5) is temporally preserved following the flow, as we see below.

### *2.3 Lagrangian quantum hydrodynamics*

We have shown how the spacetime trajectories defined by the integral curves of the velocity field (the trajectories employed by de Broglie and Bohm in their hidden-variable interpretation) may be invoked to demonstrate propagation properties of a quantum system. We now demonstrate that, subject to suitable initial conditions, they specify the quantum dynamics completely. To proceed, we observe that the single-valuedness of the velocity field implies that the initial position coordinates $q_{0i}$ uniquely specify the trajectories. They therefore provide a continuously variable set of three labels to identify the curves, which we denote $q_{0i} = a_i$. The full congruence is therefore described by the displacement function $q_i(a,t)$, which is single-valued and differentiable with respect to $a_i$ and $t$ and the inverse mapping $a_i(q,t)$ exists and has the same properties. We may thus conceive of the system as comprising a continuum of fluid elements, or 'particles', the identity of each being preserved throughout the flow and defined by the invariant $a_i$. This step is not merely of mathematical significance for the labeling allows us to conceive of fluid functions such as density and pressure in terms of notions not available in the Eulerian picture, namely, interparticle interactions described by the deformation matrix $\partial q_i / \partial a_l$. This description of the state, using a system of coordinates moving with the medium, corresponds to the Lagrangian picture of a fluid.

Suppose we have some means of calculating the trajectories that does not depend on first knowing the velocity field. Then, according to a result due to Euler [6], they generate the general solution for any fields $\rho, v_i$ that satisfy the continuity equation (2.7). Consider the following identity obeyed by the microscopic particle density $\delta(x - q(a,t))$:

$$\frac{\partial}{\partial t}\delta(x-q(a,t)) + \frac{\partial}{\partial x_i}\left[\delta(x-q(a,t))\frac{\partial q_i(a,t)}{\partial t}\right] = 0. \qquad (2.14)$$

Multiplying by a function $\rho_0(a)$ and summing over all $a_i$ we obtain the following formulas for functions $\rho, v_i$ obeying the continuity equation in terms of the functions $q_i(a,t)$:

$$\rho(x,t) = \int \rho_0(a)\delta(x-q(a,t))d^3a \qquad (2.15)$$

$$\rho(x,t)v_i(x,t) = \int \rho_0(a)\frac{\partial q_i(a,t)}{\partial t}\delta(x-q(a,t))d^3a. \qquad (2.16)$$

As suggested by the notation, $\rho_0 = |\psi_0|^2$ is the initial value of $\rho$. Computing the integrals we get the local relations

$$\rho(x,t) = J^{-1}\big|_{a(x,t)} \rho_0(a(x,t)) \qquad (2.17)$$

$$v_i(x,t) = \frac{\partial q_i(a,t)}{\partial t}\bigg|_{a(x,t)} \qquad (2.18)$$

where

$$J = \det(\partial q/\partial a) = \frac{1}{3!}\varepsilon_{ijk}\varepsilon_{lmn}\frac{\partial q_i}{\partial a_l}\frac{\partial q_j}{\partial a_m}\frac{\partial q_k}{\partial a_n}, \quad 0 < J < \infty. \tag{2.19}$$

The relations (2.17) and (2.18) give the desired general solution. Note that in the trajectory language the continuity equation (2.7) becomes the conservation law

$$\rho(q,t)d^3q(a,t) = \rho_0(a)d^3a. \tag{2.20}$$

Since $\rho$ is the particle number density, (2.20) states that the number of particles contained in an elementary volume is conserved by the flow. This relation prompts defining the mass of a fluid element as $m\rho d^3q$ since, assuming $\rho$ is normalized, the total mass of the fluid is $m$.

To complete the trajectory representation of the fluid functions we need a self-contained dynamical equation to determine the vector function $q_i$. This follows immediately from the Euler equation (2.8). From (2.18) we deduce the following relation between the accelerations in the two pictures:

$$\frac{\partial v_i}{\partial t} + v_j\frac{\partial v_i}{\partial x_j} = \left.\frac{\partial^2 q_i(a,t)}{\partial t^2}\right|_{a(x,t)}. \tag{2.21}$$

Substituting $x_i = q_i(a,t)$, Euler's force law then becomes in the Lagrangian picture

$$m\frac{\partial^2 q_i(a)}{\partial t^2} = -\frac{\partial}{\partial q_i}\left(V(x)+V_Q(x)\right)\bigg|_{x=q(a,t)}. \tag{2.22}$$

Here derivatives with respect to $q_i$ are regarded as shorthand for derivatives with respect to $a_i$ via the formula

$$\frac{\partial}{\partial q_i} = J^{-1}J_{ij}\frac{\partial}{\partial a_j} \tag{2.23}$$

where $J_{ij}$ is the adjoint of the deformation matrix $\partial q_i/\partial a_l$ with

$$\frac{\partial q_i}{\partial a_j}J_{il} = J\delta_{lj}, \quad J_{il} = \frac{\partial J}{\partial(\partial q_i/\partial a_l)} \tag{2.24}$$

and $\rho$ in $V_Q$ is given by (2.17). The initial condition corresponding to (2.11) is $\partial q_{i0}(a)/\partial t = m^{-1}\partial S_0(a)/\partial a$ and the dynamics is completed by specifying $\rho_0(a)$, which appears explicitly in (2.22) via $V_Q$.

We conclude that (2.22) is *Schrödinger's equation in the form of Newton's second law*. The quantum state is now represented by the 'displacement amplitude'

$q_i(a,t)$ encoding the history of an infinite ensemble of particles whose interaction is described by the derivatives of $q_i$ with respect to $a_i$ (which appear up to fourth order on the right-hand side of (2.22)). With the appropriate initial conditions the vector $q_i(a)$ determines the motion completely, without reference to $\psi(x)$. Complementary to the latter's unitary evolution, *quantum evolution is represented as the deterministic unfolding of a continuous coordinate transformation* $a_i \to q_i$. We remark that one can reverse the demonstration and deduce Schrödinger's equation starting from Newton's law, as shown previously [1].

As an application of this form of Schrödinger's equation we easily derive a quantum version of Kelvin's theorem on the conservation of circulation [12,13]:

$$\frac{\partial}{\partial t}\oint \dot{q}_i\, dq_i = 0. \qquad (2.25)$$

Here the closed loop, composed of particles, remains closed during the flow due to the continuity of the function $q_i(a)$. The symmetry corresponding to this conservation law is the covariance of the theory with respect to continuous transformations of the particle label [14].

Starting from the $\psi$-representative of the quantum state we may compute the $q_i$-representative by solving (2.18). Conversely, starting from the solutions $q_i$ we may evaluate $\psi$ using the following prescription. First of all, the initial data $\rho_0(a)$, $\dot{q}_{oi}$ fixes the initial wavefunction $\psi_0(a) = \sqrt{\rho_0}\exp(iS_0/\hbar)$ up to an irrelevant constant phase. To compute the wavefunction for all $x,t$ up to a global phase we first solve (2.22) subject to the initial conditions $q_{0i}(a) = a_i$, $\partial q_{i0}(a)/\partial t = m^{-1}\partial S_0(a)/\partial a_i$ to get the set of trajectories for all $a_i, t$. Next, substitute $q_i(a,t)$ in (2.17) to find $\rho$ and $\partial q_i/\partial t$ in (2.18) to get $m^{-1}\partial S/\partial x_i$. This gives $S$ up to an additive function of time, $f(t)$. To fix this function, apart from an additive constant, use (2.3). We obtain finally the following formula for the wavefunction as a function of the trajectory solution:

$$\psi(x,t) = \sqrt{\left(J^{-1}\rho_0\right)\big|_{a(x,t)}}\exp\left[\frac{i}{\hbar}\left(\int m\,\partial q_i(a,t)/\partial t\big|_{a(x,t)}\,dx_i + f(t)\right)\right]. \qquad (2.26)$$

To summarize, we have presented three forms of the Schrödinger equation and associated concepts of state and examined their equivalence: the usual one (2.1) of wave mechanics where the state is represented by $\psi(x)$, the Eulerian hydrodynamic version with state $\rho(x), v_i(x)$, and the Lagrangian hydrodynamic version with state $q_i(a)$, the initial velocity in the last two being subject to the gradient condition.

We remark that the single-body theory presented here readily extends to an $n$-body system by allowing the indices $i,j,\ldots$ to range over $3n$ values [1,2]. The configuration space congruence may be mapped into ensembles of interlacing trajectories in 3-space. Generalizing the configuration space to a Riemannian manifold the method of representing the quantum state by Lagrangian coordinates embraces a wide variety of systems including spin ½ and fields [2,15,16] and other implications have been explored [17,18]. A related remark is that the change of variables through which the Schrödinger equation is recast in Newtonian form is not unique. For example, relativistic considerations suggest an alternative expression for the non-relativistic velocity [19] and other models exist [20] but we shall not pursue these here.

### 3. Hamiltonian formulation of quantum mechanics

It is well known that, using Hamiltonian methods, the Schrödinger equation may be expressed in a form that closely mirrors classical mechanics [21,22] although the interpretations given to the symbols representing the physical state in the respective phase spaces are radically different (probability amplitude and conjugate momentum *vs*. particle position and conjugate momentum). Here we shall examine some aspects of the usual Hamiltonian version of quantum mechanics. In the following two sections we show how the quantum phase space variables may be chosen to resemble more closely the classical state.

We recall that in classical mechanics the state of a physical system is described by conjugate variables $q_i, p_i, i=1,2,3$, obeying Hamilton's equations

$$\dot{q}_i = \frac{\partial H_{qp}}{\partial p_i}, \quad \dot{p}_i = -\frac{\partial H_{qp}}{\partial q_i}. \tag{3.1}$$

In the following, continuous transformations of the variables $q_i, p_i, H_{qp} \to q'_i, p'_i, H_{q'p'}$ that leave Hamilton's equation invariant – the canonical transformations – will be particularly significant. There are various necessary and sufficient signatures that a mapping is canonical. For example, if the generating function depends on the old coordinates and new momenta, $W(q, p')$, so that the transformation equations are

$$p_i = \frac{\partial W}{\partial q_i}, \quad q'_i = \frac{\partial W}{\partial p'_i}, \quad H_{q'p'} = H_{qp} + \frac{\partial W}{\partial t}, \tag{3.2}$$

one characterization of canonicity is the invertibility of the Hessian matrix $h_{ij} = \partial^2 W / \partial q_i \partial p'_i$. An alternative characterization is the invariance of the Poisson brackets under the transformation.

To see the close analogy between the classical and quantum Hamiltonian formalisms it is useful to reformat Hamilton's equations in complex coordinates. Thus, introducing the coordinates $z_i = (q_i + ip_i)/\sqrt{2}$ with conjugate momenta $\pi_i = iz_i^*$, (3.1) become

$$\dot{z}_i = \frac{\partial H_{z\pi}}{\partial \pi_i}, \quad \dot{\pi}_i = -\frac{\partial H_{z\pi}}{\partial z_i} \tag{3.3}$$

where $H_{z\pi} = H_{qp}$. This transformation from real to complex phase space is evidently canonical. Extending the range of $i$, Schrödinger's equation and its complex conjugate have just this form for a discrete quantum system if $z_i$ is identified with complex normal coordinates and the Hamiltonian describes a collection of oscillators [21].

The Hamiltonian formulation of the Schrödinger equation in the position representation may be obtained by extending the discrete complex version of

Hamilton's equations to a continuous system. In the usual variational approach the Hamiltonian is written [23]

$$H_{\psi\pi}[\psi,\pi] = (i\hbar)^{-1} \int \left( \frac{\hbar^2}{2m} \frac{\partial \pi}{\partial x_i} \frac{\partial \psi}{\partial x_i} + V(x)\pi\psi \right) d^3x \tag{3.4}$$

where $\psi$ and $\pi = i\hbar\psi^*$ are conjugate variables. This is the mean value of the Hamiltonian operator in the state $\psi$: $H_{\psi\pi} = \langle \psi | \hat{H} | \psi \rangle$. Hamilton's equations,

$$\dot{\psi} = \frac{\delta H_{\psi\pi}}{\delta \pi}, \quad \dot{\pi} = -\frac{\delta H_{\psi\pi}}{\delta \psi}, \tag{3.5}$$

reproduce the Schrödinger equation (2.1) and its complex conjugate. Here the functional derivative is defined as follows [24]. Suppose $F$ is a functional of some function $\phi(x)$: $F[\phi] = \int f(x,\phi,\partial\phi,\partial^2\phi,...) d^3x$. Then the functional derivative of $F$ with respect to $\phi$ is

$$\frac{\delta F}{\delta \phi} = \frac{\partial f}{\partial \phi} - \frac{\partial}{\partial x_i} \frac{\partial f}{\partial(\partial\phi/\partial x_i)} + \frac{\partial^2}{\partial x_i \partial x_j} \frac{\partial f}{\partial(\partial^2\phi/\partial x_i \partial x_j)} - \cdots \tag{3.6}$$

Using this notation the functional Hamilton equations may be expressed in terms of Poisson brackets, the latter being defined for two functionals $A, B$ of phase space variables $\phi(y), \gamma(y)$ as follows:

$$\{A,B\}_{\phi\gamma} = \int \left( \frac{\delta A}{\delta \phi(y)} \frac{\delta B}{\delta \gamma(y)} - \frac{\delta B}{\delta \phi(y)} \frac{\delta A}{\delta \gamma(y)} \right) d^3y. \tag{3.7}$$

Being a Hamiltonian system, the Schrödinger equation admits canonical transformations as symmetries, which link the old phase space coordinates $\psi(x), \pi(x)$ with a new set $\psi'(y), \pi'(y)$ that obey Hamilton's equations with a transformed Hamiltonian $H_{\psi'\pi'}$. Unitary transformations $U(y,x)$ form a class of canonical transformations where the old and new coordinates are connected linearly, and likewise for the momenta:

$$\psi'(y) = \int U(y,x)\psi(x) d^3x, \quad \pi'(y) = \int U^*(y,x)\pi(x) d^3x \tag{3.8}$$

with

$$\int U(y,x)U^*(y,x') d^3y = \delta(x-x'), \quad \int U(y,x)U^*(y',x) d^3x = \delta(y-y'). \tag{3.9}$$

Since the old and new coordinates are functionally related we shall assume that the generating functional of the canonical transformation representing the unitary transformation is a functional of the old coordinates and new momenta:

$$W[\psi,\pi'] = \int \pi'(y) U(y,x) \psi(x) d^3y\, d^3x. \tag{3.10}$$

The equations of the canonical transformation, continuous analogues of (3.2), are

$$\psi'(y) = \frac{\delta W}{\delta \pi'(y)} = \int U(y,x) \psi(x) d^3x, \quad \pi(x) = \frac{\delta W}{\delta \psi(x)} = \int \pi'(y) U(y,x) d^3y, \tag{3.11}$$

which, using (3.9), reproduce (3.8). The canonical character of the transformation is confirmed by evaluating the Hessian matrix,

$$\frac{\delta^2 W}{\delta \pi'(y) \delta \psi(x)} = U(y,x), \tag{3.12}$$

which is invertible by (3.9). The transformed Hamilton equations with new Hamiltonian

$$H_{\psi'\pi'} = H_{\psi\pi} + \frac{\partial W}{\partial t} \tag{3.13}$$

where $H_{\psi'\pi'} = \langle \psi' | \hat{H}' | \psi' \rangle$, $\hat{H}' = \hat{U}\hat{H}\hat{U}^\dagger + i\hbar (d\hat{U}/dt) \hat{U}^\dagger$, yield Schrödinger's equation and its complex conjugate for the new variables $\psi'(y)$. Hence, the quantum-mechanical transformation theory can be fully incorporated into the Hamiltonian language.

Non-unitary canonical transformations also arise in quantum mechanics and as a simple example we observe that the passage to the polar representation of the wavefunction is a transformation of this type. Let the functions $\psi, \pi$ ($\rho, S$) be the old (new) coordinates and momenta, respectively. Assuming the generating functional depends on the old and new coordinates $\psi, \rho$, we choose

$$W[\psi,\rho] = i\hbar \int \rho \left[ \log(\psi/\sqrt{\rho}) + 1/2 \right] d^3x. \tag{3.14}$$

Then, from the equations of a canonical transformation,

$$\pi = \frac{\delta W}{\delta \psi}, \quad S = -\frac{\delta W}{\delta \rho}, \tag{3.15}$$

we deduce the following explicit formulas for the old variables in terms of the new:

$$\psi = \sqrt{\rho} e^{iS/\hbar}, \quad \pi = i\hbar \sqrt{\rho} e^{-iS/\hbar}. \tag{3.16}$$

The Hessian matrix

$$h(x,x') = \frac{\delta^2 W}{\delta \rho(x') \delta \psi(x)} = \frac{i\hbar}{2\psi(x)} \delta(x-x') \tag{3.17}$$

has inverse $h^{-1}(x,x') = (2\psi(x)/i\hbar)\delta(x-x')$, showing that the mapping is indeed canonical. We also easily confirm the invariance of the Poisson brackets (3.7) in passing between the two sets of phase space coordinates:

$$\{A,B\}_{\psi\pi} = \{A,B\}_{\rho S}. \tag{3.18}$$

The Hamiltonian is a scalar under this time-independent transformation and its new form is

$$H_{\rho S}[\rho,S] = \int \rho\left[\frac{1}{2m}\frac{\partial S}{\partial x_i}\frac{\partial S}{\partial x_i} + U(\rho,\partial\rho) + V\right]d^3x \tag{3.19}$$

where

$$U = \frac{\hbar^2}{8m}\frac{1}{\rho^2}\frac{\partial\rho}{\partial x_i}\frac{\partial\rho}{\partial x_i} \tag{3.20}$$

is the internal quantum potential energy. Hamilton's equations,

$$\dot{\rho} = \frac{\delta H_{\rho S}}{\delta S}, \quad \dot{S} = -\frac{\delta H_{\rho S}}{\delta \rho}, \tag{3.21}$$

reproduce (2.2) and (2.3) where the quantum potential (2.4) is obtained here via the formula

$$V_Q(x) = \frac{\delta}{\delta\rho(x)}\int \rho(x')U(\rho(x'))d^3x'. \tag{3.22}$$

### 4. Canonical formulation of the Schrödinger equation using density and current density potentials

The canonical theory just presented is economical in that no extraneous variables beyond those of physical interest ($\psi,\pi$ or $\rho,S$) appear but it is not suitable for setting up a canonical relation with the trajectory theory. For this purpose an enhanced phase space is required. Fortunately, there are other ways to formulate the Schrödinger equation in Hamiltonian terms. Noting that the Eulerian-picture hydrodynamic formulation of the Schrödinger equation provided a fruitful intermediary between the wave-mechanical and trajectory theories in Sec. 2, our strategy will be to first seek a canonical formulation of that picture of quantum hydrodynamics. Abetted by this, we shall set up a canonical transformation to the Lagrangian trajectory model (next section).

We start by writing the Hamiltonian (3.19) in terms of the variables $\rho, v_i$:

$$H_{\rho v} = \int \left(\tfrac{1}{2}m\rho v_i v_i + \rho U(\rho,\partial\rho) + \rho V\right)d^3x. \tag{4.1}$$

For the present we consider general vortical flows, not confined by the constraint (2.6). The expression (4.1) for the energy formally falls within the scope of classical hydrodynamics, the principal difference with typical classical fluids being the specific derivative form of the internal quantum energy $U$. It is a well-known feature of fluid mechanics that the Eulerian-picture variables $\rho, v_i$ are not canonical [25]. To achieve a canonical formulation requires introducing potentials for the velocity, which results in a Clebsch-like representation: $v_i = \chi_j \partial_i \lambda_j$ (as indeed arises naturally in the $\rho$, $S$-based approach of the previous section). It is customary to employ three Clebsch parameters ($v_i = \partial_i \theta + \chi \partial_i \lambda$) but, if they are single-valued, this is not sufficient to represent the most general vector field [26,27] (three parameters can suffice, however, if multivalued functions are admitted [28]). In our development below an extended Clebsch-like representation (with six parameters) arises naturally. In the first instance this is introduced as a representation of the mass current density $m\rho v_i$ rather than $v_i$ directly.

A straightforward way to achieve the required parameterization is to make a space- and time-dependent transformation of the space coordinates to new independent variables: $x_i \rightarrow a_i = Q_i(x,t)$. Under this transformation the dependent variables $\rho$ (number density) and $m\rho v_i$ (mass current density = momentum density) transform according to the usual formulas of a coordinate substitution for tensor densities [29]:

$$\rho(x,t) = j(Q(x,t))\rho_0(Q(x,t)) \qquad (4.2)$$

$$m\rho(x,t)v_i(x,t) = \overline{P}_j(Q(x),t) j(Q(x,t)) \frac{\partial Q_j(x,t)}{\partial x_i}$$
$$= -P_j(x,t) \frac{\partial Q_j(x,t)}{\partial x_i} \qquad (4.3)$$

where

$$j = \det(\partial Q/\partial x) = \frac{1}{3!}\varepsilon_{ijk}\varepsilon_{lmn}\frac{\partial Q_i}{\partial x_l}\frac{\partial Q_j}{\partial x_m}\frac{\partial Q_k}{\partial x_n}, \quad 0 < j < \infty. \qquad (4.4)$$

Here the transformed momentum density is denoted $\overline{P}_i$ and we have written $\overline{P}_i j$ as the function $-P_i(x)$. The minus sign in the latter definition is introduced for later convenience (in particular, so that a '+' sign appears in (4.11) below). Following Sec. 2 the mapping is assumed to be single-valued and differentiable and the inverse mapping $a_i \rightarrow x_i = q_i(a,t)$ exists and has the same properties. The transformation must obey two additional conditions. A first requirement is that the transformed density is the initial function $\rho_0$, expressed in the $Q$-coordinates, and hence has no explicit time dependence. This is necessary so that $\rho$ obeys the continuity equation in the $x$-coordinates (see below). The second condition, which puts physical content into the transformation, is that the new dependent variables $Q_i, P_i$ are canonically conjugate, i.e., they define three pairs of position and momentum variables for each space point whose temporal development is governed by Hamilton's equations. Mathematically, the new variables replace $\rho, v_i$ as fundamental descriptors of the state and in terms of them the Hamiltonian (4.1) becomes

$$H_{QP}[Q,P] = \int \left( \frac{1}{2m\rho} P_i P_j \frac{\partial Q_i}{\partial x_k} \frac{\partial Q_j}{\partial x_k} + \rho U(\rho, \partial \rho) + \rho V \right) d^3x \quad (4.5)$$

where $\rho$ is shorthand for the function (4.2) and $U$ is given in (3.20). The physical interpretation of the functions $Q_i$ will be considered in the next section.

The functions $Q_i$ in (4.2) evidently provide a 'density vector potential' since, given $\rho_0$, we may determine the physical density $\rho$ by differentiation. These functions also appear as potentials for the mass current (4.3). We thus obtain a Clebsch-type representation of the velocity $v_i$ with six parameters $Q_i$ and $-P_i/m\rho$.

The Clebsch parameters corresponding to a given vector field are not unique and equally viable sets $\chi_i, \lambda_i$ and $\chi'_i, \lambda'_i$ are connected by a canonical transformation [1,30,31]. In our case this 'gauge' freedom in the parameters $Q_i, P_i$ is restricted (but still represented by a canonical transformation) because we have the novel feature that the subset $Q_i$ determines $\rho$ as well as $v_i$. To obtain the general form of the gauge transformation, suppose that the physical fields, the left-hand sides in (4.2) and (4.3), are connected with a different set of independent canonical variables $Q'_i, P'_i$ according to the same formulas. Then the transformation linking the two canonical sets $Q_i, P_i$ and $Q'_i, P'_i$ that leaves the space coordinates and physical fields invariant is easily seen to be a time-independent difffeomorphism $Q'_i = f_i(Q)$ with

$$\left. \begin{array}{l} x'_i = x_i, \quad \rho' = \rho, \quad v'_i = v_i, \quad P'_i(x) = P_j(x) \dfrac{\partial Q_j}{\partial f_i} \\ \rho'_0(Q'(x,t)) = \det(\partial f / \partial Q) \rho_0(Q(x,t)). \end{array} \right\} \quad (4.6)$$

That these relations define a canonical transformation may be verified by deriving them using the formulas

$$W[Q,P'] = \int f_i(Q(x)) P'_i(x) d^3x, \quad Q'_i = \frac{\delta W}{\delta P'_i}, \quad P_i = \frac{\delta W}{\delta Q_i}, \quad H_{Q'P'} = H_{QP}. \quad (4.7)$$

Below we shall exploit the gauge freedom to simplify the theory.

To confirm that $Q_i, P_i$ are suitable canonical variables, we examine Hamilton's equations:

$$\frac{\partial Q_i}{\partial t} = \frac{\delta H_{QP}}{\delta P_i} = \frac{1}{m\rho} P_k \frac{\partial Q_k}{\partial x_j} \frac{\partial Q_i}{\partial x_j} \quad (4.8)$$

$$\frac{\partial P_i}{\partial t} = -\frac{\delta H_{QP}}{\delta Q_i} = j^{-1} j_{ij} \left[ \frac{\partial}{\partial x_k} \left( \frac{1}{m\rho} P_m P_n \frac{\partial Q_m}{\partial x_k} \frac{\partial Q_n}{\partial x_j} \right) \right.$$
$$\left. - P_n \frac{\partial}{\partial x_j} \left( \frac{1}{m\rho} P_m \frac{\partial Q_m}{\partial x_k} \frac{\partial Q_n}{\partial x_k} \right) + \rho \frac{\partial}{\partial x_j} (V + V_Q) \right] \quad (4.9)$$

where $j$ is given by (4.4), $j_{ij}$ is the adjoint of the inverse deformation matrix $\partial Q_i/\partial x_l$ with

$$\frac{\partial Q_i}{\partial x_j} j_{il} = j\delta_{lj}, \quad j_{il} = \partial j/\partial(\partial Q_i/\partial x_l), \tag{4.10}$$

$V_Q$ is given in (2.4) and $\rho$ in (4.2). These equations are unfamiliar but their physical content is easily revealed. Using the expression (4.3) for $v_i$, the first equation (4.8) states that the vector function $Q_i$ is constant following the flow generated by $v_i$:

$$\frac{\partial Q_i}{\partial t} + v_j \frac{\partial Q_i}{\partial x_j} = 0. \tag{4.11}$$

This is how conservation is represented in these variables. In accordance with $Q_i$'s role as a vector potential for the density (4.2), the continuity equation (2.7) may be derived by differentiating (4.11) with respect to $x_j$. Thus, using the formulas (4.4) and

$$\varepsilon_{ijk} \frac{\partial Q_i}{\partial x_p} = j\varepsilon_{prs} \frac{\partial x_r}{\partial Q_j} \frac{\partial x_s}{\partial Q_k} \tag{4.12}$$

we obtain

$$\frac{\partial j}{\partial t} + \frac{\partial}{\partial x_i}(jv_i) = 0. \tag{4.13}$$

Eq. (4.11) ensures that $\rho_0(Q)$ is a constant of the motion and hence (2.7) follows from (4.13). With the aid of the continuity equation it may be shown that the second Hamilton equation (4.9) is equivalent to the quantum Euler force law (2.8). Hamilton's equations therefore imply the correct hydrodynamic equations (2.7) and (2.8).

To establish equivalence with Schrödinger's equation, we must fix the initial conditions $Q_{0i}(x), P_{0i}(x)$. Following (4.2) and (4.3) the latter are connected with the initial hydrodynamic functions according to

$$\rho(x,t=0) = j(Q_0(x))\rho_0(Q_0(x)), \quad m\rho(x,t=0)v_{0i}(x) = -P_{0k}(x)\frac{\partial Q_{0k}}{\partial x_i}. \tag{4.14}$$

These relations determine $Q_{0i}, P_{0i}$ up to a gauge transformation (4.6). To simplify matters we fix the gauge so that $Q_{0i}(x) = x_i$ since this is the condition for which $\rho(x,t=0) = \rho_0(x)$. Invoking the initial gradient form (2.11) of $v_i$ we then have, altogether,

$$Q_{0i}(x) = x_i, \quad P_{0k}(x) = -\rho_0(x)\partial_i S_0(x), \quad \rho_0 = |\psi_0|^2. \tag{4.15}$$

To summarize, Hamilton's equations (4.8) and (4.9) with initial conditions (4.15) imply the Schrödinger equation via the quantum hydrodynamic equations (2.7) and (2.8). Conversely, we may deduce Hamilton's equations from the Schrödinger equation on substituting the relations (4.2) and (4.3) for $\rho$ and $v_i$ into the quantum hydrodynamic equations. We conclude that the phase space variables $Q_i, P_i$ provide a satisfactory alternative canonical formulation of quantum propagation.

Using (4.3), (4.10) and (4.11) we can solve for $P_i$ in terms of $\dot{Q}_i$:

$$P_i = m\rho g_{ij} \frac{\partial Q_j}{\partial t} \tag{4.16}$$

where $g_{ij} = j^{-2} j_{ik} j_{jk}$ and $\rho = j(Q)\rho_0(Q)$. Substituting this expression in (4.9) we may eliminate $P_i$ to obtain a self-governing second-order equation for the gauge potentials $Q_i$, which now represent the quantum state. This is the Eulerian picture equivalent of the second-order Lagrangian picture equation (2.22) (to which it is related via a canonical transformation, as we shall see below). This second-order version of Schrödinger's equation (which we do not give explicitly) may also be written as an Euler-Lagrange equation once the Lagrangian is obtained. The latter is found by making an inverse Legendre transformation of the Hamiltonian (4.5) and then substituting expression (4.16) for $P_i$:

$$\begin{aligned} L[Q,\dot{Q},t] &= \int P_i(x)\dot{Q}_i(x)d^3x - H_{QP}[Q,P,t] \\ &= \int \rho \left( \frac{1}{2} m g_{ij}(x) \frac{\partial Q_i(x)}{\partial t} \frac{\partial Q_j(x)}{\partial t} - \frac{\hbar^2}{8m} \frac{\partial \log \rho}{\partial x_i} \frac{\partial \log \rho}{\partial x_i} - V(x) \right) d^3x \end{aligned} \tag{4.17}$$

with $g_{ij} = j^{-2} j_{ik} j_{jk}$ and $\rho = j(Q)\rho_0(Q)$.

We note that to compute $Q_i$ from a known velocity field $v_i$ we may use (4.11). This is the Eulerian equivalent of using the Lagrangian equation (2.18) to solve for $q_i$ given $v_i$.

The relation between the first-order ($\psi$ or $\rho, v_i$) and second-order ($Q_i$) formulations of Schrödinger's equation is analogous to that between the first-order (electric+magnetic fields) and second-order (vector potential) versions of Maxwell's equations, each formulation being self-contained with respect to its dependent variables with the two sets of initial conditions being chosen to ensure compatibility. As with electromagnetism, the quantum potentials-based approach exhibits a gauge symmetry and provides a variational basis for the theory. The quantum approach parts company with electromagnetism in one key respect, however: in the quantum case it is possible to formulate the theory using gauge-invariant potentials. These potentials are just the displacement functions $q_i$, as we see next.

## 5. Canonical transformation to the spatial trajectory formulation

With the canonical formulation of the last section at hand, it is straightforward to transform the Schrödinger equation into Newtonian form by performing a canonical transformation $Q_i(x), P_i(x) \to q_i(a), p_i(a)$ whose remit is to effect an inversion of the independent and dependent variables, $x_i \leftrightarrow a_i$. We suppose that the generating functional of the transformation depends on the new coordinates and old momenta and is time-independent: $W[q(a), P(x)]$. The transformation formulas are therefore

$$Q_i(x) = \frac{\delta W}{\delta P_i(x)}, \quad p_i(a) = \frac{\delta W}{\delta q_i(a)}, \quad H_{qp}[q,p,t] = H_{QP}[Q,P,t]. \tag{5.1}$$

A generating functional with the required inversion property is given by

$$W[q,P] = \int \delta(x - q(a)) J a_i P_i(x) d^3x d^3a \tag{5.2}$$

with $J$ defined in (2.19). For

$$Q_i(x) = \frac{\delta W}{\delta P_i(x)} = \int \delta(x - q(a)) J a_i d^3a \tag{5.3}$$

and using the formula $\delta(x - q(a)) = J^{-1} \delta(a - q^{-1}(x))$ gives $Q_i(x) = q_i^{-1}(x)$. The other transformation formula yields

$$p_i(a) = \frac{\delta W}{\delta q_i(a)} = -J(q(a)) \frac{\partial a_j(q(a))}{\partial q_i(a)} P_j(q(a)) \tag{5.4}$$

where we have used the result

$$\frac{\delta J(a')}{\delta q_i(a)} = -J_{ij}(a) \frac{\partial}{\partial a_j} \delta(a - a') \tag{5.5}$$

and the identity $\partial J_{ij}/\partial a_j = 0$ with $J_{ij}$ defined in (2.24). The explicit solution for the new phase space variables in terms of the old is thus

$$q_i(a,t) = a_i^{-1}(x,t) \tag{5.6}$$

$$p_i(a,t) = -J(x) P_j(x) \frac{\partial Q_j(x)}{\partial x_i} \bigg|_{x=q(a)} = m \rho_0(a) v_i(q(a,t), t). \tag{5.7}$$

The Hessian for the transformation is

$$h_{ij}(x,a) = \frac{\delta W}{\delta P_j(x) \delta q_i(a)} = -\frac{\partial a_j(q(a))}{\partial q_i(a)} \delta(a - Q(x)) \tag{5.8}$$

with inverse

$$h_{ij}^{-1}(x,a) = -\frac{\partial q_i(a)}{\partial a_j}\delta(x-q(a)), \qquad (5.9)$$

which confirms the canonical nature of the mapping. This is confirmed also by the invariance of the Poisson brackets:

$$\{A,B\}_{QP} = \{A,B\}_{qp} \qquad (5.10)$$

The new Hamiltonian is

$$H_{qp}[q,p,t] = \int \left[ \frac{1}{2m\rho_0(a)} p_i(a)p_i(a) + \rho_0(a)U(J^{-1}\rho_0) + \rho_0(a)V(q(a)) \right] d^3a \qquad (5.11)$$

and in these variables Hamilton's equations take a recognisable form:

$$\frac{\partial q_i(a)}{\partial t} = \frac{\delta H_{qp}}{\delta p_i(a)} = p_i(a)/m\rho_0(a) \qquad (5.12)$$

$$\frac{\partial p_i(a)}{\partial t} = -\frac{\delta H_{qp}}{\delta q_i(a)} = -\rho_0(a)\frac{\partial}{\partial q_i}(V+V_Q) \qquad (5.13)$$

where $\partial/\partial q_i$ is given in (2.23). Eliminating $p_i$ from these equations we obtain the second-order Newtonian equation (2.22) for $q_i$. To complete the canonical mapping we state the initial conditions obeyed by the new variables corresponding to the gauge used in (4.15) for the old variables:

$$q_{0i}(a) = a_i, \quad p_{0i}(a) = \rho_0(a)\partial_i S_0(a), \quad \rho_0 = |\psi_0|^2. \qquad (5.14)$$

As with the coordinates $Q_i$ in (4.2) and (4.3), the coordinates $q_i$ in (2.17) and (2.18) provide a set of potentials from which the hydrodynamic functions are obtained by differentiation: $\rho = \det(\partial q/\partial a)^{-1}\rho_0$, $v_i = \partial q_i/\partial t$. The gauge transformation (4.6) translates here into a time-independent difffeomorphism $a'_i = f_i(a)$, or relabeling of the trajectories, with respect to which

$$\left.\begin{array}{l} q'_i(a',t) = q_i(a,t), \quad p'_i(a',t) = \det(\partial f/\partial a) p_i(a,t), \\ \rho'_0(a') = \det(\partial f/\partial a)\rho_0(a). \end{array}\right\} \qquad (5.15)$$

As anticipated above, we see that the potentials $q_i$ are gauge invariant functions, as are the velocity components $\dot{q}_i$. Note that if we made a different choice for $q_{0i}(a) \neq a_i$, $\rho_0(a) \neq |\psi_0(a)|$.

The passage to the new variables clarifies the significance of the Eulerian-picture potentials $Q_i$: they provide a spacetime representation of the labels of the

Lagrangian-picture trajectories and their conservation expressed in (4.11) represents the invariance of the particle label $a_i$ along the line of flow it characterizes.

Transforming to the $q_i$-coordinates, the Lagrangian (4.17) becomes

$$L[q,\dot{q},t] = \int \rho_0(a)\left[\frac{1}{2}m\frac{\partial q_i(a)}{\partial t}\frac{\partial q_i(a)}{\partial t} - \frac{\hbar^2}{8m}G_{ij}\frac{\partial \log\rho}{\partial a_i}\frac{\partial \log\rho}{\partial a_j} - V(q(a))\right]d^3a \quad (5.16)$$

where $G_{ij} = J^{-2}J_{ki}J_{kj}$ and $\rho = J^{-1}(a)\rho_0(a)$. The Euler-Lagrange equation gives (2.22). It will be noted how the metric characterizing the deformation appears in the kinetic term in the $Q$-representation ($g_{ij}$ in (4.17)) and in the potential term in the $q$-representation ($G_{ij}$ in (5.16)). The canonical theory was developed previously starting from the Lagrangian (5.16) [1].

We conclude once again that the deterministic continuous trajectory is tacitly contained in wave mechanics rather than being an additional structure: *The trajectory model of quantum evolution is obtained by a canonical transformation of wave mechanics when this is formulated in terms of the hydrodynamic phase space variables, or gauge potentials, $Q_i(x), P_i(x)$.*

## 6. The quantum formalism in *q,p* phase space

We saw in Sec. 3 that quantum dynamics may be formulated as a set of Hamilton equations where the Hamiltonian is the mean value of the Hamiltonian operator. We show here how the mean value, regarded as a functional, may be employed more generally to represent the quantum operator calculus and use this result to interpret the latter in terms of the alternative phase space representation *q,p* of the quantum state.

According to the usual formalism, to any observable represented by a self-adjoint operator $\hat{A}(x',x)$ (in the position representation) we may associate a real-valued bilinear functional of the conjugate variables $\psi, \pi$ via its mean value:

$$A[\psi,\pi] = \langle\psi|\hat{A}|\psi\rangle = (i\hbar)^{-1}\int \pi(x')\hat{A}(x',x)\psi(x)d^3x\,d^3x'. \quad (6.1)$$

This is the matrix representation of the operator in the quantum phase space spanned by $\psi, \pi$ and we may work with the functional $A$ instead of the operator $\hat{A}$. The explicit form of the latter is obtained by differentiation:

$$\hat{A}(x',x) = i\hbar\frac{\delta^2 A[\psi,\pi]}{\delta\pi(x')\delta\psi(x)}. \quad (6.2)$$

The operator algebra may be translated into the functional calculus of mean values as follows. Introducing a second operator $\hat{B}$ with associated functional $B$ defined as in (6.1), the sum and product of $\hat{A}, \hat{B}$ have the following associated functionals:

$$\langle\psi|\hat{A}+\hat{B}|\psi\rangle = A[\psi,\pi] + B[\psi,\pi] \quad (6.3)$$

$$\langle\psi|\hat{A}\hat{B}|\psi\rangle = i\hbar \int \frac{\delta A}{\delta\psi(x)} \frac{\delta B}{\delta\pi(x)} d^3x. \tag{6.4}$$

Using these formulas we may compute the mean value of an arbitrary function of operators from the mean values of its component operators. An immediate application of this result is the well known [21] theorem that the mean value of $((i\hbar)^{-1}$ times) the commutator $[\hat{A},\hat{B}] = \hat{A}\hat{B} - \hat{B}\hat{A}$ is the Poisson bracket of the functionals $A,B$ in the phase space $\psi,\pi$:

$$\langle\psi|(i\hbar)^{-1}[\hat{A},\hat{B}]|\psi\rangle = \{A,B\}_{\psi\pi} = \int\left(\frac{\delta A}{\delta\psi(x)}\frac{\delta B}{\delta\pi(x)} - \frac{\delta B}{\delta\psi(x)}\frac{\delta A}{\delta\pi(x)}\right)d^3x. \tag{6.5}$$

Having represented the operator calculus in the $\psi,\pi$ phase space, insight into its physical content may be obtained by passing to the $q,p$ phase space. Following (2.26) and (5.7) we can translate (6.1) into these variables as follows. For the wavefunction we write, ignoring the phase factor $f(t)$,

$$\psi(x=q(a,t),t) = \sqrt{(J^{-1}(a,t)\rho_0(a))}\exp\left(\frac{i}{\hbar}\int \frac{p_i(a,t)}{\rho_0(a)}\frac{\partial q_i}{\partial a_j}da_j\right). \tag{6.6}$$

The conjugate momentum $\pi(x=q(a,t),t)$ is given by $i\hbar$ times the complex conjugate of (6.6). For $x_i$ and $\partial/\partial x_i$ in the algebraic and differential function $\hat{A}$ we write $q_i$ and $\partial/\partial q_i = J^{-1}J_{ij}\partial/\partial a_j$, respectively, and in integrals $d^3x$ is replaced by $Jd^3a$. This implies for (6.1) a formula of the type

$$A[q,p] = \int f(q(a),p(a),q'(a'),p'(a'),\partial q(a),\partial q'(a'),...)d^3a\,d^3a'. \tag{6.7}$$

We can express the operator (6.2) and the composition formula (6.4) in terms of $q,p$ by writing the functional derivatives of the latter in terms of those for $\psi,\pi$. A basic result is that the mean value of the commutator (6.5) may be expressed as the Poisson brackets for the new variables when the functionals $A,B$ depend on $q,p$ through $\psi,\pi$:

$$\int\left(\frac{\delta A}{\delta q_i(a)}\frac{\delta B}{\delta p_i(a)} - \frac{\delta B}{\delta q_i(a)}\frac{\delta A}{\delta p_i(a)}\right)d^3a = \int\left(\frac{\delta A}{\delta\psi(x)}\frac{\delta B}{\delta\pi(x)} - \frac{\delta B}{\delta\psi(x)}\frac{\delta A}{\delta\pi(x)}\right)d^3x. \tag{6.8}$$

An efficient way to prove this result is to express the functional derivatives on the left-hand side in terms of the polar variables,

$$\frac{\delta}{\delta q_i(a)} = \int \left( \frac{\delta S(x)}{\delta q_i(a)} \frac{\delta}{\delta S(x)} + \frac{\delta \rho(x)}{\delta q_i(a)} \frac{\delta}{\delta \rho(x)} \right) d^3x,$$

$$\frac{\delta}{\delta p_i(a)} = \int \left( \frac{\delta S(x)}{\delta p_i(a)} \frac{\delta}{\delta S(x)} + \frac{\delta \rho(x)}{\delta p_i(a)} \frac{\delta}{\delta \rho(x)} \right) d^3x,$$

(6.9)

and invoke the Poisson bracket equivalence (3.18). The result (6.8) is a quantum analogue of a special case of the general relation between classical Lagrangian and Eulerian (non-Poisson) brackets [32].

We conclude that we may transcribe the quantum operator calculus into the new variables via the intermediary of mean values. For the basic set of variables of interest in physics the formula (6.7) yields simple and readily interpretable expressions:

position
$$\langle \hat{x}_i \rangle = \int x_i \rho(x,t) d^3x = \int q_i(a,t) \rho_0(a) d^3a \qquad (6.10)$$

linear momentum
$$\langle \hat{p}_i \rangle = \int \partial_i S(x,t) \rho(x,t) d^3x = \int p_i(a,t) d^3a \qquad (6.11)$$

angular momentum
$$\langle (\hat{x} \times \hat{p})_i \rangle = \int (x \times \partial S)_i \rho d^3x = \int (q \times p)_i d^3a \qquad (6.12)$$

kinetic energy
$$\left\langle \frac{1}{2m} \hat{p}_i \hat{p}_i \right\rangle = \int \left( \frac{1}{2m} \rho \partial_i S \partial_i S + \rho U(\rho) \right) d^3x = \int \left( \frac{p_i p_i}{2m\rho_0} + \rho_0 U(J^{-1}\rho_0) \right) d^3a. \qquad (6.13)$$

Note that this rendering of quantum theory differs fundamentally from phase space approaches such as that of Wigner where $q,p$ are the independent variables and the state is represented by a quasi-distribution function $f(q,p)$ (constructed from the wavefunction) whose evolution is governed by a quantum analogue of the classical Liouville equation. In our case the phase space variables define the state and the independent variables specify the particles, an analogue of classical fluid theory in its Lagrangian phase space formulation. There is, however, one property the Wigner approach and ours share: neither is unique. A wide class of quasi-distribution functions may be employed to represent the quantum state, and likewise for the law of motion in the fluid model.

## 7. Observability of the quantum state

We have seen that the hydrodynamic state variables $\rho, v_i$ play a key role in connecting the $\psi$- and $q_i$-representations of the state. We now show how these functions may also be employed to derive empirical information about the state. Assuming a means to observe these functions has been found, we may deduce the $\psi$, $q_i$-versions by integration as follows:

1. $\psi(x,t)$: From $\rho$ we deduce $|\psi|$ and from $v_i = m^{-1} \partial_i S$ we integrate to get the phase $S$ up to an additive function of $t$ that is fixed apart from a constant by substituting in (2.3).

2. $q_i(a,t)$: From $v_i$ we get $q_i(a)$ by integrating $v_i = \dot{q}_i$ with $a_i = q_{0i}$.

In seeking empirical methods to investigate the state, we first exclude one potential avenue by noting that the velocity field $v_i$ is not a quantum observable: there does not exist a linear Hermitian operator for which $v_i$ is its expectation value [33-35]. On the other hand, the local density $\rho$ and current density $j_i$ ($=\rho v_i$) *are* observables in the quantum-mechanical sense: there exist Hermitian operators $\hat{\rho}_x, \hat{j}_{x_i}$ such that $\rho(x) = <\hat{\rho}_x>$, $j_i(x) = <\hat{j}_{x_i}>$ for all $\psi$. Specifically,

$$\hat{\rho}_x = |x_i\rangle\langle x_i|, \quad \hat{j}_{x_i} = \frac{1}{2m}\left[|x_i\rangle\langle x_i|\hat{p}_i + \hat{p}_i^\dagger|x_i\rangle\langle x_i|\right], \quad \text{(no sum over } i\text{)}. \quad (7.1)$$

The 'non-observability' of $v_i$ may be proved by applying the following result [34] (for a comment on the proof see [35]):

*Theorem*: Consider three non-trivial (i.e., not multiples of the identity operator) operators $\hat{A}, \hat{B}, \hat{C}$ where $\hat{A}, \hat{B}$ are linear and the mean values obey the relation $<\hat{A}> = <\hat{B}><\hat{C}>$ for all states in Hilbert space. Then $\hat{C}$ cannot be linear.

To apply the theorem choose $\hat{A} = \hat{j}_{x_i}, \hat{B} = \hat{\rho}_x$ for each $i = 1,2,3$. Then the ratio $v_i = j_i/\rho$ is not the mean value of a linear operator. A slightly different version of the theorem with the same implication for $v_i$ is proved in the Appendix.

Of course, as a spacetime function $v_i$ *is* 'observable' as it may be deduced statistically from measurements of quantum observables over an ensemble of identically prepared systems (same $\psi_0$). For example, one can perform a sequence of position and momentum measurements at each spacetime point, which yields the local velocity [3]. Such statistical methods may be used both to learn the state when it is unknown and to check the predictions it encodes when it is known through some prior state preparation process. The methods do not, however, impinge on the ontological status of the state, in particular the hydrodynamic functions. To address the ontological issue we consider the possibility of a direct observation of the hydrodynamic functions using the protective measurement procedure.

With the aim of demonstrating an ontological aspect of the wavefunction, Aharonov and co-workers [36-38] (for clarifications and reviews see [39,40]) showed how a suitably adapted adiabatic interaction described by quantum mechanics provides a scheme to measure the expectation values of operators pertaining to a system without appreciably disturbing its quantum state. These interactions are therefore called 'protective measurements'. Aharonov *et al.* claimed that in certain circumstances this technique 'measures the wavefunction' of a single system as an extended object. They infer that the procedure, in revealing a property possessed by a single system prior to the measurement, provides evidence for the ontological character of the wavefunction. In fact, we shall see that it is the hydrodynamic functions that are measured by this method rather than the wavefunction directly.

Consider two interacting systems, an object and measuring apparatus, with initial wavefunctions $\psi_0(x)$ and $\beta_0(y)$, respectively. We assume for simplicity that the configuration coordinates *x* and *y* are one-dimensional. Denote by $\hat{A}$ the operator pertaining to the object whose expectation value is to be measured. Then in a protective interaction the total Hamiltonian comprises free Hamiltonians for the individual systems and an interaction term $\hat{H}_I = g(t)y\hat{A}$. The initial combined state $\Phi_0(x,y) = \psi_0(x)\beta_0(y)$ then evolves adiabatically at time *t* into:

$$\Phi(x,y,t) = \psi(x,t)\beta(y,t)\exp\left[-(i/\hbar)\int_T^t g(t)y\langle\hat{A}\rangle dt\right]. \quad (7.2)$$

Here *g(t)* is a function characterizing the adiabatic interaction with $\int_T^t g(t)dt = 1$, and $\psi(x,t)$ and $\beta(y,t)$ are the wavefunctions obtained under free evolution of the two systems. It will be observed that (7.2) is still a product state in that the variables *x* and *y* have not become entangled. In particular, the object state is undisturbed by the interaction. The state of the apparatus has, however, acquired a phase factor depending on the expectation value $\langle\hat{A}\rangle = \langle\psi(t)|\hat{A}|\psi(t)\rangle$, which implies an observable change in the apparatus momentum. Hence, information on the state $\psi(x,t)$ can be gleaned from the apparatus by measuring the change in its momentum (via a conventional measurement). The method may be extended to protectively measure several operators simultaneously by introducing corresponding additional apparatuses. Thus, for a second operator $\hat{B}$ we introduce an apparatus with coordinate *z* and wavefunction $\gamma(z)$ and the formula (7.2) becomes

$$\Phi(x,y,z,t) = \psi(x,t)\beta(y,t)\gamma(z,t)\exp\left[-(i/\hbar)\int_0^t \left(g_y(t)y\langle\hat{A}\rangle + g_z(t)z\langle\hat{B}\rangle\right)dt\right]. \quad (7.3)$$

Let us choose $\hat{A} = \hat{j}_{x_i}, \hat{B} = \hat{\rho}_x$ for each *i* so that $\langle\hat{A}\rangle = \rho(x,t)v_i(x,t), \langle\hat{B}\rangle = \rho(x,t)$. Then the momenta of the devices *y* and *z* are shifted by the amounts $\int_0^t g_y(t)\rho(x,t)v_i(x,t)dt$ and $\int_0^t g_z(t)\rho(x,t)dt$, respectively.

We thus have a scheme to measure the time-averaged density and current density at a space point *x*. Aharonov *et al*. based their claim of 'measuring the wavefunction of a single system' on these formulas, for the special case where $\psi$ is known to be a non-degenerate energy eigenstate but is otherwise unknown. For in that case the density and current density are time-independent and the momentum shifts are proportional to the local values. For this particular case we may therefore measure the functions $\rho, v_i$ for all values of their arguments and deduce $\psi$ (up to a gauge transformation) as noted in 1 above. As already remarked, it is the hydrodynamic variables that are measured in this scheme; $\psi$ is deduced from them.

The protective method may be applied to states other than non-degenerate energy eigenstates but there are two caveats: (a) the full Hamiltonian that functions during the protective process depends on the state [37], which implies that we must first know $\psi$ before we can investigate it, and (b) that investigation reveals results about time averages of (the hydrodynamic) functions of $\psi$ rather than instantaneous values. So, in the general case the protective scheme provides a way to empirically confirm time-averaged prior information about the state.

We conclude that functions of the hydrodynamic fields associated with a single quantum system are measurable quantities, namely, their time-averaged local values for a general state that is known prior to the measuring procedure. As a special case, the fields themselves are measurable when the state is an energy eigenstate but otherwise unknown. In the latter case the quantum state in its $\psi$ or $q_i$ guise may be deduced as a single entity from the empirical data, as set out above. An important consideration in relation to the $q_i$-state is the consistency of this result with the unfeasibility of simultaneously measuring position and momentum, which would require the $\psi$-state to transform into a simultaneous eigenstate of the associated operators. The latter impossibility appears to be consistent with the protective measurement scheme since $q_i$ is a construct from its results rather than the object of investigation. For reasons discussed elsewhere [3], the simultaneous attribution of position and momentum variables to each particle $a_i$ is also consistent with the uncertainty relations, which constitute conditions on the statistical scatter of position and momentum measurement results in accord with the quantum formalism.

## 8. A common language for quantum and classical physics?

We have seen that the self-contained second-order Newtonian version of the Schrödinger equation,

$$m\frac{\partial^2 q_i(a)}{\partial t^2} = -J^{-1}J_{ij}\frac{\partial}{\partial a_j}\left(V(x)+V_Q(x)\right)\bigg|_{x=q(a,t)} \qquad (8.1)$$

subject to $\rho_0(a)=|\psi_0(a)|^2$, $\dot{q}_{oi}(a)=m^{-1}\partial S_0(a)/\partial a_i$, attributes quantum evolution to the motion of a continuous infinity of interacting particles (in the fluid sense) having simultaneously well-defined position and momentum variables. This approach therefore appears to supply the basis of a common language for quantum and classical mechanics and might be expected to provide insight into how the theories are connected. As we shall see, while the theory does assist in this aim, it also highlights several subtleties if the quest is to treat classical-like behaviour as a limiting case of an enveloping quantum description, something that is almost universally regarded as desirable and feasible.

Following (8.1), an obvious criterion for a quantum system to behave like a classical one, at least approximately, is that the quantum contribution to the force, $\partial_i V_Q$, is negligible compared to the classical force, for this rendering of Schrödinger's equation then reduces to Newton's classical law. This correspondence principle is state-dependent, which has the advantage of encompassing and explaining the fact that procedures which seek to characterize the limit in terms of the relative values of parameters such as $m$ and $\hbar$ often do not lead to classical behaviour [3, 41]. However, the state-dependent limiting process stated in this form is not sufficient to characterize the classical domain for several reasons.

As we have noted, a congruence of trajectories is needed for a complete description of the quantum state and none is singled out for special status, beyond the selection implicit in the initial conditions for the density and velocity. In addition, in quantum mechanics the probability density $\rho$ refers to the likely position should a measurement of location be performed and not to the probability of current presence. The limiting process envisaged above maintains both the congruence and the quantum version of probability; it does not yield the single trajectory expected for a classical particle or the classical statistical concept based on the actual presence of the particle independent of measurement. These problems are not insurmountable and may be addressed by invoking the postulates of the Broglie-Bohm theory [3]: that one of the paths $a_i$ in the collective representing the state supports a material corpuscle, and $\rho$ refers to its likely current position. Then the ensemble of paths representing the quantum state may be viewed as the potential paths of the added corpuscle, only one of which is actually realized with a frequency determined by $\rho$.

But even if this step is taken there remains a further problem with the Newton-based limiting protocol [41]: the correspondence limit is exceptional. As noted above, classical behaviour does not generally emerge when parametrical limiting processes are applied to specific wavefunctions (quantum systems with no classical analogue). But even where a quantum state is found that does imply the classical domain in some limit, the classical behaviour so obtained may be but a subset of that which is allowed by classical laws for that system (classical systems with no quantum analogue). A corollary is that quantum wavefunctions cannot generally solve typical classical statistical problems. These points may be illustrated with simple examples, such as reflection of a particle by a wall [41].

The fundamental impediment to obtaining classical motion according to the scheme based on (8.1) is the congruent character of the state: at each moment at most one trajectory passes through each space point. When the putative limiting procedures are taken, the single-valuedness condition remains intact and constrains each *individual* orbit so that the set of all individuals obey the non-crossing property. This is not a characteristic of generic classical particle ensembles for which there is no single-valuedness requirement. The same problem would, however, occur in classical fluid mechanics if we wanted to consider circumstances in which the internal fluid forces are negligible relative to external body forces. In fact, we expect this would not generally be possible in a real fluid because it would destroy the mechanism that maintains its physical integrity. Is demanding a comparable reduction in the effectiveness of the internal forces in the quantum case likewise too stringent a requirement?

We make three observations about these issues. The first is that our analysis pertains to pure states and it has been suggested that the difficulties might be forestalled if the limiting process is treated using mixed states [42].

The second remark is that the issues raised may be artefacts of the position representation that go away in a different formulation of quantum mechanics (for an alternative approach see [43]).

The third point is that our considerations refer to closed systems. In practice, unless special conditions are available to isolate a system, macroscopic bodies are generally in continual interaction with their environment. It has been suggested in connection with the so-called 'decoherence' programme (for a review see [44]) that the 'classicality' of a system may be contingent on these background interactions. This mechanism may also be invoked in the context of the de Broglie-Bohm theory as was first suggested in connection with a specific problem: that of obtaining the classical motion of a planet using the ambient stellar light as the environmental entity whose action causes potentially interfering classical segments of the planet wavefront to 'decohere' [45] (another component of the problem, showing how a resultant Keplerian orbit is obtained for the planet, was not addressed in this work). To see what is involved, suppose we have a set of wavefunctions $\psi_\mu$ each of which is 'classical' in the sense of negligible associated quantum force. Then, in a region where the functions overlap, the total wavefunction is given by $\sum_\mu \psi_\mu$ and the particle motion in the region of superposition is generally non-classical since the associated finite quantum force will act to preserve the congruent character of the trajectories. To disrupt this interference effect, suppose the system interacts with another system having many degrees of freedom and wavefunction $\phi$. Then, for a suitable interaction Hamiltonian, each wave $\psi_\mu$ will couple to $\phi$ in a different way with a resultant total wavefunction $\sum_\mu \psi_\mu \phi_\mu$. For a sufficiently complex external system each summand $\psi_\mu \phi_\mu$ occupies a distinct region of the configuration space and the de Broglie-Bohm configuration point will lie in and be guided by just one summand. Thus the subsystem of interest will be guided by just one of the waves $\psi_\mu$ (while the others remain finite) and its classicality is ensured.

To explain classical behavior in general, the decoherence process would have to supply an interaction Hamiltonian that spontaneously, constantly and widely acts in nature to bring about persistently non-overlapping configuration space packets. This means that classical mechanics, insofar as it can be discerned as a special case, emerges as a sub-dynamics in the total configuration space of the trajectory theory corresponding to the system-plus-environment. In this way the decoherence and trajectory theories may be mutually supportive in that each brings elements that the other lacks but further analysis is needed (see for example [46]).

## 9. Conclusion

We have sought to show that Newton's trajectory law involving a particular type of interaction potential (the quantum potential) and subject to suitable initial conditions is just Schrödinger's equation expressed in different variables. In particular, this formulation may be obtained by a canonical transformation of wave mechanics. The displacement concept of state may thus be regarded as implicit in the quantum description. In the case of classical continuum physics the transformation we have described – between the field-theoretic Eulerian version and the particle-like Lagrangian version – is unexceptional since the subject starts from a Newtonian analysis of physical systems as extended to continua [6]. The state of the system naturally has complementary forms. In the quantum case the analogous result acquires a more potent significance since the theory is supposed to depart so radically from classical notions that it is felt that a trajectory formulation, even if possible, must be at best a metaphysical supplement of no import to the theory itself. To be sure, the physical significance of the quantum trajectories, such as what may travel along them, may be an issue of interpretation but their definition and employment as an alternative and independent characterization of the quantum state is a matter of mathematical transformation.

According to these results the notion of 'wave-particle duality' has some currency but it refers to different ways of viewing a single process, not to mutually exclusive experimental contexts as has been asserted historically. Our thesis developed elsewhere [2,17,18] is that this duality is not specific to quantum theory (or to classical continuum theories) but is an aspect of generic field theories that may be expected to admit Lagrangian trajectory formulations.

Is this alternative version of quantum mechanics useful? It is already well established that a trajectory outlook brings computational benefits to quantum theory [4]. It also allows us to examine questions that may be difficult to formulate in the $\psi$-description, such as criteria for chaos (e.g., [47]), time of transit (e.g., [48]) and, as we have seen, classical-like behaviour. This is a field where a great deal of foundational and numerical work has been done but it still lacks significant theorems. It also lacks empirical support although, as we have seen, a scheme may be conceived to measure the hydrodynamic functions with which the trajectory variables are intimately connected. Perhaps the observation that, in the end, the trajectory is just a different way of doing quantum mechanics may attract further interest.

## Appendix: Proof that the velocity field is not a quantum observable

*Theorem*: Consider three operators $\hat{A}, \hat{B}, \hat{C}$ where $\hat{A}, \hat{B}$ are linear and the mean values ($A_\psi = \langle \psi | \hat{A} | \psi \rangle$ etc.) obey the relation

$$A_\psi = B_\psi C_\psi \tag{H.2}$$

for all Hilbert space states $|\psi\rangle$ with $A_\psi, B_\psi \neq 0$ for at least one state. Then $\hat{C}$ cannot be linear.

*Proof*: We may use (H.2) and the linearity of $\hat{A}, \hat{B}$ to compute $A_{2\psi}$ in two ways:

$$A_{2\psi} = 4A_\psi = 4B_\psi C_\psi \tag{H.3}$$

$$A_{2\psi} = B_{2\psi} C_{2\psi} = 4B_\psi C_{2\psi}. \tag{H.4}$$

Equating the right-hand sides of these relations we have $C_{2\psi} = C_\psi$. If $\hat{C}$ is linear this implies $C_\psi = 0$, which contradicts (H.2) if $|\psi\rangle$ is a state for which $A_\psi, B_\psi \neq 0$. Hence $\hat{C}$ cannot be linear. □

As noted in the text, to apply the theorem put $\hat{A} = \hat{j}_{x_i}$, $\hat{B} = \hat{\rho}_x$ for each $i = 1, 2, 3$. Then the velocity component $v_i = j_i / \rho$ is not the mean value of a linear operator. Note that this result holds whatever definition is adopted for the density and current density fields, so long as these are expressible as mean values of linear operators. In particular, it applies to alternative trajectory theories such as that derived from the non-relativistic limit of the Dirac theory where the current density necessarily differs from that used here [19]. The latter example provides a salutary lesson that a conserved current derived from a wave equation may contain a trivial (identically conserved) component that is physically non-trivial.